\documentclass[a4paper,11pt]{article}
\usepackage{pos}

\usepackage{epsf,amsmath,graphicx,dcolumn}
\usepackage{scalefnt}
\usepackage{hyperref}
\usepackage{cleveref,etoolbox,color,soul}
\usepackage{listings,booktabs}
\usepackage[utf8]{inputenc}
\usepackage{epsfig}
\usepackage{graphicx}
\usepackage[colorinlistoftodos]{todonotes}
\usepackage{multirow}
\usepackage{bbold}

\def\Y{\mathbf{Y}}

\def\Mnu{\mathbf{M}_\nu}
\def\Mnuh{\widehat{\mathbf{M}}_\nu}

\def\U{\mathbf{U}}

\def\dmatm{\Delta m^2_{31}}
\def\dmsol{\Delta m^2_{21}}

\def\EW{$\mathrm{SU(2)}_L \otimes \mathrm{U(1)}_Y$}
\newcommand{\BR}{{\rm BR}}

\newcolumntype{K}[1]{>{\centering\arraybackslash}m{#1}}
\def\gsim{\raise0.3ex\hbox{$\;>$\kern-0.75em\raise-1.1ex\hbox{$\sim\;$}}}
\def\lsim{\raise0.3ex\hbox{$\;<$\kern-0.75em\raise-1.1ex\hbox{$\sim\;$}}}

\title{Hybrid scoto/seesaw: flavour and dark matter}

\author[a]{D.~M. Barreiros}
\author*[a]{H.~B. C\^amara}
\author[a]{F.R.~Joaquim}

\affiliation[a]{Departamento de F\'{\i}sica and CFTP, Instituto Superior T\'ecnico, Universidade de Lisboa, Av. Rovisco Pais 1, 1049-001 Lisboa, Portugal}

\emailAdd{henrique.b.camara@tecnico.ulisboa.pt}
\emailAdd{debora.barreiros@tecnico.ulisboa.pt}
\emailAdd{filipe.joaquim@tecnico.ulisboa.pt}

\abstract{We propose a model based on the interplay between the type-II seesaw and scotogenic neutrino mass generation mechanisms. The setup features a $\mathbb{Z}_8$ discrete flavour symmetry which is broken down to a residual $\mathbb{Z}_2$ responsible for stabilising dark matter. A singlet scalar field is introduced to implement spontaneous CP violation. The effective neutrino mass matrix two-texture zero structure leads to sharp neutrino sector predictions. We analyse the constraints imposed on the model by current and future charged lepton flavour violation experiments. This framework provides two viable dark matter candidates, scalar or fermion. We investigate the scalar dark matter scenario considering relic density, direct-detection and collider constraints.}

\FullConference{%
  8th Symposium on Prospects in the Physics of Discrete Symmetries (DISCRETE 2022)\\
  7-11 November, 2022\\
  Baden-Baden, Germany
}

\begin{document}
\maketitle

\section{Introduction}

The Standard Model (SM) of particle physics cannot explain neutrino flavour oscillations which imply massive neutrinos and lepton mixing, nor the observed dark matter (DM) relic abundance. From a theoretical perspective tackling these problems requires going beyond the SM. Arguably, the simplest possible extension of the SM, that accommodates neutrino masses, relies on the addition of a single Higgs triplet with hypercharge $+1$, i.e. the type-II seesaw mechanism~\cite{Schechter:1980gr,Mohapatra:1980yp}. Additionally, one can explore scenarios where the reason neutrinos are massive is connected to DM. Namely, in the scotogenic model~\cite{Ma:2006km}, neutrino masses arise at one-loop mediated by {\em dark} messengers which can be suitable DM particles. The most economical scotogenic scenario contains two Majorana fermion singlets and an inert Higgs doublet. This {\em dark} sector is odd under a $\mathbb{Z}_2$ symmetry, stabilising the lightest odd particle and providing a fermion or scalar weakly interacting massive particle (WIMP) DM candidate. It could also be that neutrino masses receive contributions both at the tree and loop level via different mechanisms, as in the scoto-seesaw~\cite{Rojas:2018wym}. The coexistence of two distinct neutrino mass mechanisms allows for vacuum-induced leptonic CP violation~(LCPV), as shown in ref.~\cite{Barreiros:2020gxu}.

Motivated by the scoto-seesaw idea, in this work we investigate flavour and DM in a hybrid scotogenic and type-II seesaw model. As in ref.~\cite{Barreiros:2020gxu}, we implement spontaneous CP violation (SCPV) via the complex vacuum expectation value~(VEV) of a scalar singlet which breaks a $\mathbb{Z}_8$ flavour symmetry down to a \textit{dark} $\mathbb{Z}_2$. The work presented here follows closely ref.~\cite{Barreiros:2022aqu}.

\section{Scoto/type-II seesaw model}

We extend the SM with one singlet fermion $f$ and four complex scalar multiplets: two doublets $\eta_i$ ($i=1,2$), one triplet $\Delta$ and one singlet $\sigma$. We impose a $\mathbb{Z}_{8}$ flavour symmetry, which will forbid some Yukawa interactions and provide low-energy predictions for neutrino mass and mixing parameters. After spontaneous symmetry breaking~(SSB), the $\mathbb{Z}_{8}$ symmetry breaks down to a {\em dark} $\mathbb{Z}_{2}$, under which $f$ and $\eta_{1,2}$ are odd, these are the {\em dark}-particles. The field content and symmetries of the model are indicated in table~\ref{tab:model}~\footnote{Depending on the charge assignment of the SM fields, two other flavour symmetries are possible: $\mathbb{Z}_8^{e-\tau}$ and $\mathbb{Z}_8^{\mu-\tau}$. Their phenomenological implications are studied in ref.~\cite{Barreiros:2022aqu}. Here we focus on the $\mathbb{Z}_{8}^{e-\mu}$ case.}. The most general Yukawa Lagrangian is
 \begin{equation}
- \mathcal{L}_{\text{Yuk.}} = \overline{\ell_L} \Y_{\ell}\, \Phi \,e_R + \overline{\ell_L} \Y_f^1 \tilde{\eta}_1 f  + \overline{\ell_L} \Y_f^2 \tilde{\eta}_2 f  + \overline{\ell_L^c} \Y_{\Delta} i \tau_2 \Delta \ell_L+ \frac{1}{2}  y_f \,\sigma \overline{f^c} f+\text{H.c.} \;,
\label{eq:LYuk}
\end{equation}
where $\ell_L=(\nu_L\; e_L)^T$ and $e_R$ denote the SM left-handed doublet and right-handed singlet charged-lepton fields, respectively. Also, $\Phi= ( \phi^+, \phi^0)^T$, $\eta_i= ( \eta_i^+, \eta_i^0)^T$ and $\Delta = (\Delta^{+ +},\Delta^{+},\Delta^{0})$, with $\tilde{\Phi}=i\tau_2 \Phi^\ast$ and $\tilde{\eta_i}=i\tau_2 \eta_i^\ast$, being $\tau_2$ the complex Pauli matrix. We impose CP invariance making all parameters in the Lagrangian real. The vacuum configuration is:
\begin{align}
  \left< \phi^0 \right> = \frac{v}{\sqrt{2}} = 174 \ \text{GeV} \; , \; \left< \eta_{1,2}^0 \right> = 0 \; , \; \left< \Delta^0 \right> = \frac{w}{\sqrt{2}} \; , \left< \sigma \right> = \frac{u\, e^{i \theta}}{\sqrt{2}} \; ,
  \label{eq:vaccum}
\end{align}
with the hierarchy $w \ll v \lsim u$. The $\mathbb{Z}_8$ symmetry allows for $\propto (\sigma^4 + \sigma^{\ast 4})$ in the scalar potential, which provides a SCPV solution where $\left< \sigma \right>$ is complex. Note that, we add the soft-breaking term  $\propto (\sigma^2 + \sigma^{\ast 2})$, to the scalar potential, leading to an arbitrary $\theta$ phase necessary to guarantee compatibility with neutrino oscillation data (see section~\ref{sec:neutrino}). The latter also avoids the cosmological domain wall resulting from the spontaneous breaking of the $\mathbb{Z}_8$ symmetry. Via the $\sigma f f$ coupling in eq.~\eqref{eq:LYuk}, the unique source of CP violation in our model, $\left< \sigma \right>$, can in principle be successfully transmitted to the fermion sector leading to non-trivial LCPV. 

\begin{table}[t!]
	\centering
	\begin{tabular}{| K{1.3cm} | K{1.3cm} | K{2.8cm} |  K{1.9cm} |}
		\hline 
&Fields&\EW&  $\mathbb{Z}_{8}^{e-\mu} \rightarrow \mathbb{Z}_2$ \\
		\hline 
		\multirow{4}{*}{Fermions} 
&$\ell_{e L} , e_R$&($\mathbf{2}, {-1/2}),(\mathbf{1}, {-1}$)& {$1$}   $\to$  $+$  \\
&$\ell_{\mu L} , \mu_R$&($\mathbf{2}, {-1/2}),(\mathbf{1}, {-1}$)& {$\omega^6$}   $\to$  $+$  \\
&$\ell_{\tau L} , \tau_R$&($\mathbf{2}, {-1/2}),(\mathbf{1}, {-1}$)& {$\omega^2$}   $\to$  $+$ \\
&$f$&($\mathbf{1}, {0}$)& {$\omega^3$}   $\to$  $-$ \\
		\hline 
		\multirow{5}{*}{Scalars}
&$\Phi$&($\mathbf{2}, {1/2}$)&$1$ $\to$ $+$\\
&$\Delta$&($\mathbf{3}, {1}$)&$1$ $\to$ $+$\\
&$\sigma$&($\mathbf{1}, {0}$)&$\omega^2$  $\to$  $+$\\
&$\eta_{1}$&($\mathbf{2}, {1/2}$)&$\omega^3$ $\to$ $-$\\
&$\eta_{2}$&($\mathbf{2}, {1/2}$)&$\omega^5$ $\to$ $-$\\
\hline
	\end{tabular}
	\caption{Fields and transformation properties under \EW and $\mathbb{Z}_8$ symmetries, where $\omega^k = e^{ik\pi/4}$.}
	\label{tab:model} 
\end{table}
\begin{figure}[t!]
    \centering
    \includegraphics[scale=0.8]{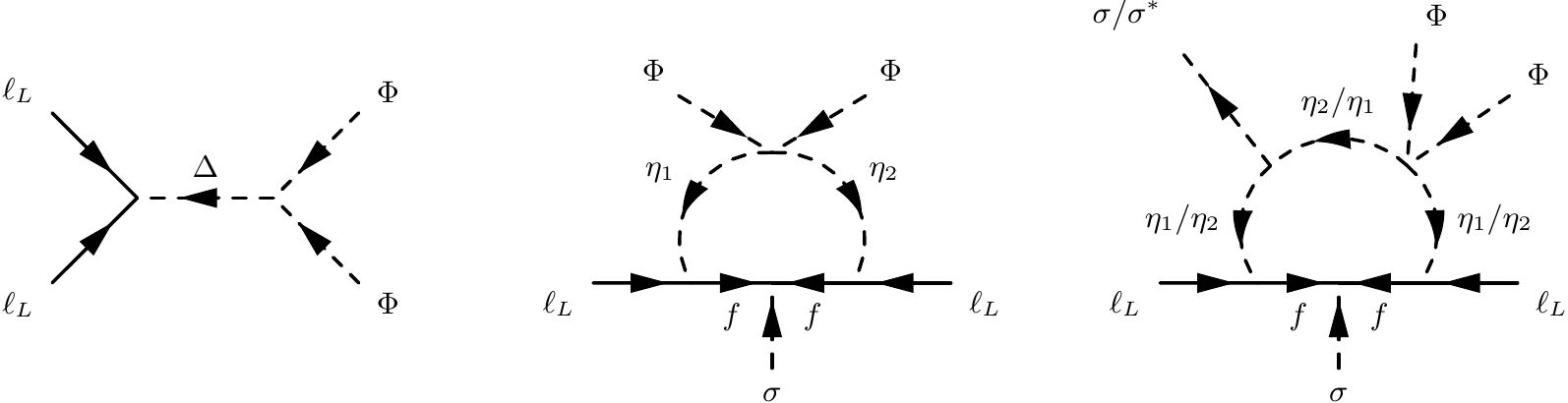}
    \caption{Type-II seesaw (left) and scotogenic (middle and right) contributions to neutrino mass generation.}
    \label{fig:diagrams}
\end{figure}
After SSB, the effective neutrino mass matrix is given by,
\begin{align}
\Mnu=\begin{pmatrix}
\mathcal{F}_{11} M_f\, y_e^2+\sqrt{2}w \,y_{1}\,e^{-i\theta} & &\mathcal{F}_{12} M_f\, y_e y_\mu& &0\\
\cdot & &\mathcal{F}_{22}M_f\, y_\mu^2 && \sqrt{2}w \,y_2e^{-i\theta}\\
\cdot & &\cdot & &0\\
\end{pmatrix}\, ,
\label{eq:Mnu}
\end{align}
where, as depicted in figure~\ref{fig:diagrams}, the terms proportional to $w/\sqrt{2}$ stem from the tree-level type-II seesaw, while the remaining terms account for the one-loop scotogenic contributions. The latter involve the loop factors $\mathcal{F}(M_f,m_{S_k})$ which depend on the $f$ mass $M_f=u\, y_f/\sqrt{2}$, and on the \textit{dark} neutral-scalar masses $m_{S_k}$ ($k=1, \cdots, 4$) and mixing resulting from the neutral components of $\eta_1$ and $\eta_2$. The two texture-zero conditions $(\Mnu)_{13}=(\Mnu)_{33}=0$ stem from the considered flavour symmetry. The interplay among neutrino mass generation mechanisms is crucial to ensure compatibility with neutrino oscillation data and non-trivial LCPV. In fact, if either $y_{e,\mu}$ or $y_{2}$ vanish, $\Mnu$ would feature an additional vanishing entry, leading to two massless neutrinos and vanishing mixing angles. Furthermore, if $y_{1}=0$, there is no CP violation in the lepton sector.

\section{Neutrino sector predictions}
\label{sec:neutrino}
\begin{figure}[!t]
    \centering
    \includegraphics[scale=0.24]{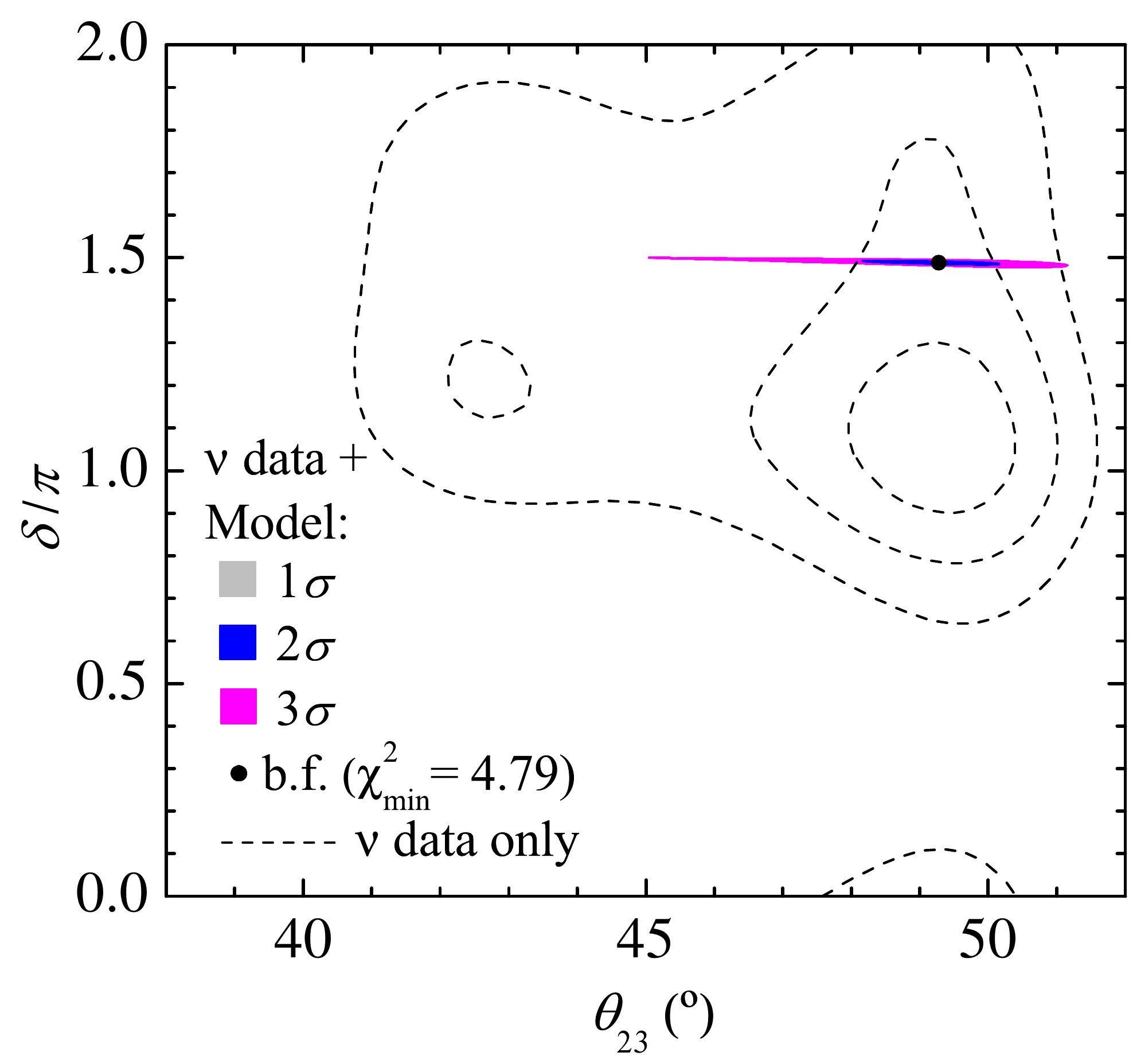} \includegraphics[scale=0.24]{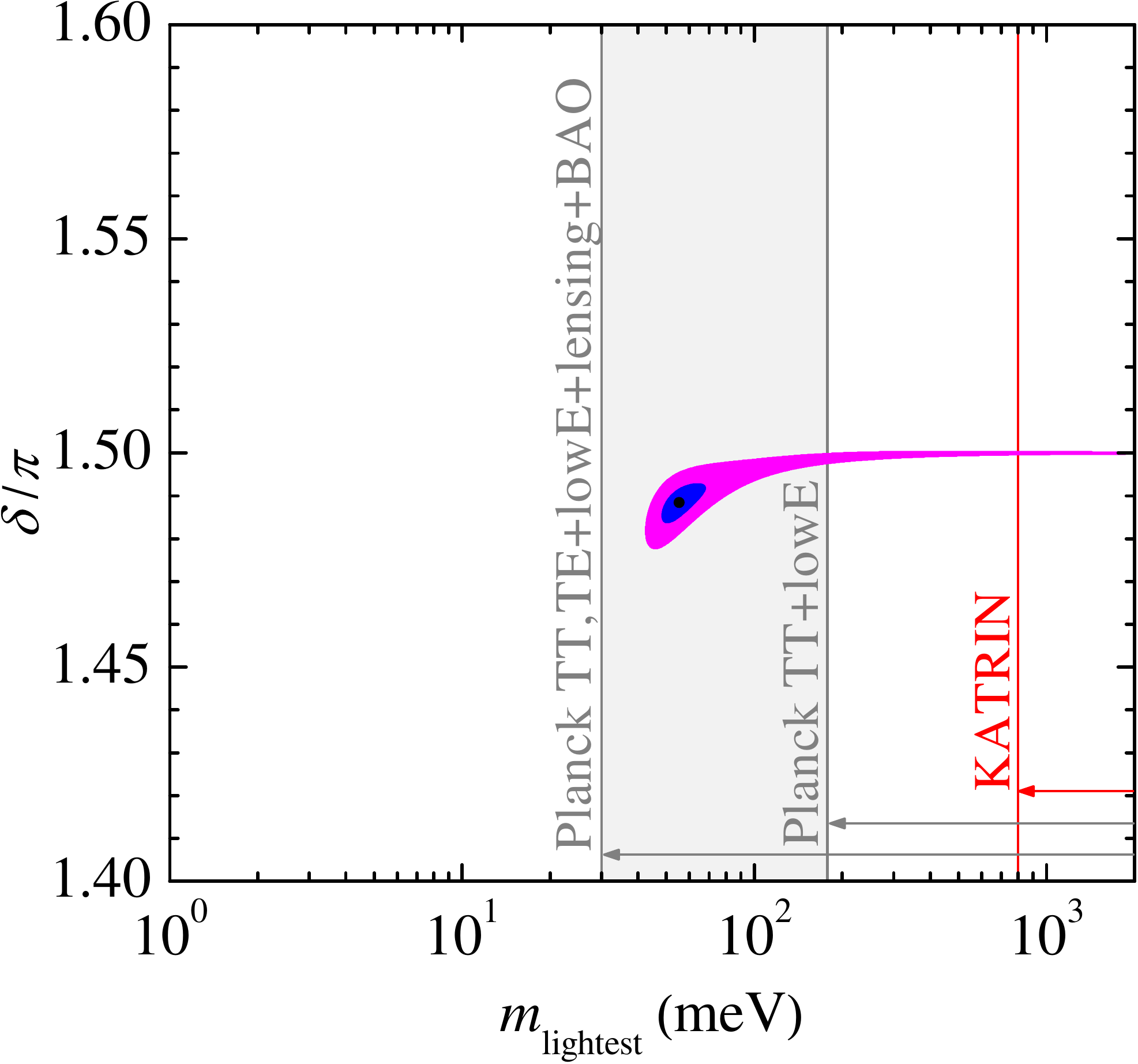} \hspace{-0.1cm}\includegraphics[scale=0.43]{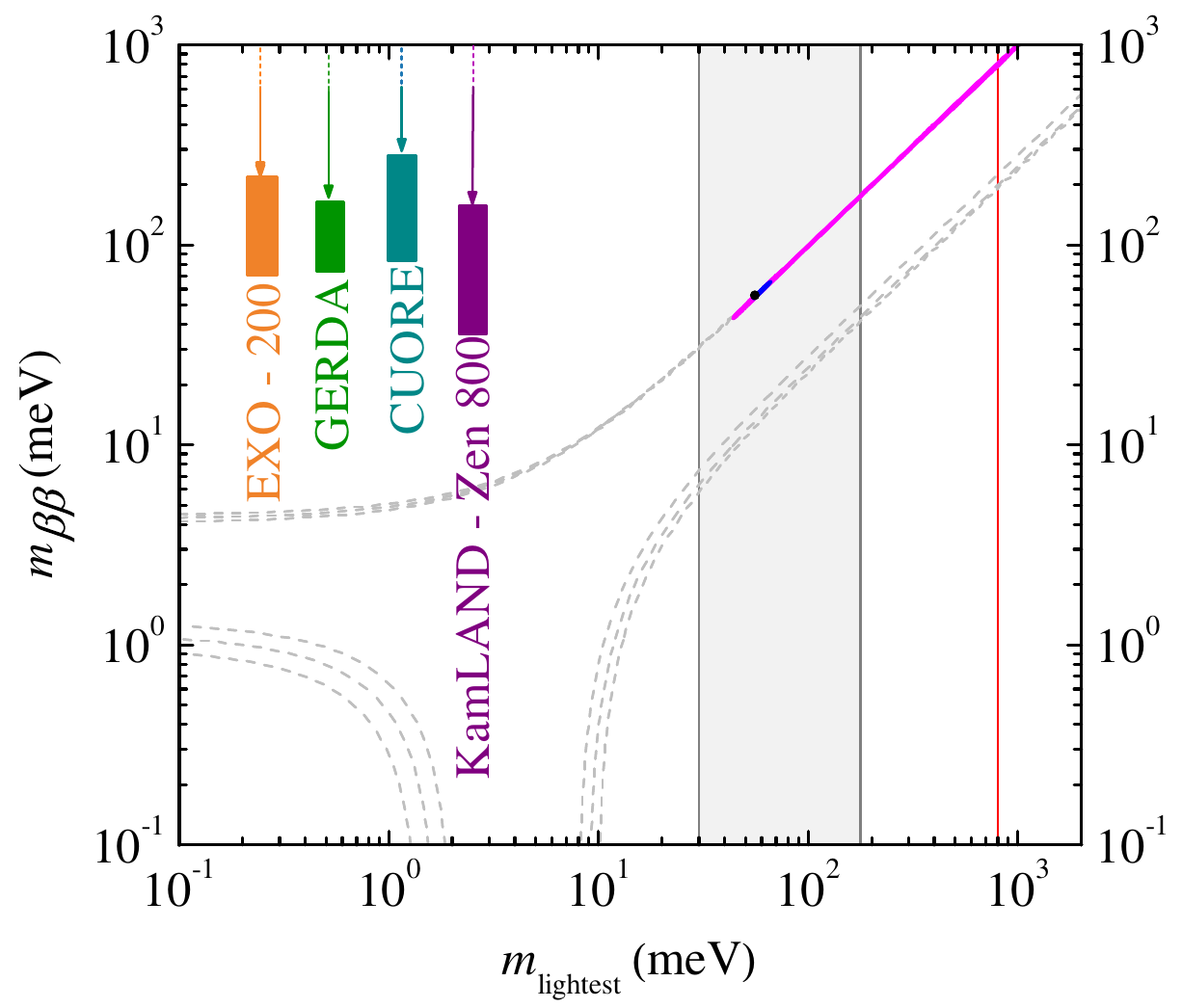}
    \caption{Allowed regions at the 1, 2 and 3$\sigma$ level (grey, blue and magenta) in the planes ($\theta_{23}$,$\delta$), ($m_\text{lightest}$,$\delta$) and ($m_\text{lightest}$,$m_{\beta\beta}$), from left to right, for NO. The black dots mark the best-fit value for each case, while the dashed contours correspond to the $\chi^2$ contours at 1, 2 and 3$\sigma$, allowed by the global fit of neutrino oscillation data~\cite{deSalas:2020pgw}. The vertical red (gray) line(s) correspond(s) to the KATRIN~\cite{KATRIN:2021uub} (Planck~\cite{Planck:2018vyg}) bound(s) on $m_\text{lightest}$. The coloured vertical bars are the current upper-bound ranges on $m_{\beta\beta}$ from the experiments looking for neutrinoless double beta decay: EXO-200~\cite{EXO-200:2019rkq}, GERDA~\cite{GERDA:2020xhi}, CUORE~\cite{CUORE:2021mvw} and KamLAND-Zen 800~\cite{KamLAND-Zen:2022tow}.}
    \label{fig:neutrino}
\end{figure}
In order to find the low-energy constraints imposed by the two-texture zero structure of eq.~\eqref{eq:Mnu}, on the neutrino-mass and mixing parameters, we first express the effective neutrino mass matrix as, $\Mnuh=\U^\ast\, \text{diag}(m_1,m_2,m_3)\,\U^\dagger$, where $m_i$ are the neutrino masses and $\U$ the lepton mixing matrix. The latter is parameterised via three mixing angles $\theta_{ij}$ ($i<j=1,2,3$), a Dirac CP-violating phase $\delta$ and two Majorana phases $\alpha_{21,31}$. Since the $\mathbb{Z}_8$ symmetry of table~\ref{tab:model} imposes a diagonal charged lepton mass matrix, the matrix $\U$ will corresponds to the one which diagonalises $\Mnu$. Neutrino oscillation data allows for two possible ordering of neutrino masses: normal and inverted neutrino-mass ordering (NO and IO). Two of the three neutrino masses may be written in terms of the lightest neutrino mass and the measured neutrino mass-squared differences $\dmsol=m_2^2-m_1^2$ and $\dmatm=m_3^2-m_1^2$. For NO we have, $m_1= m_{\rm lightest}$ and $m_2^2=m_1^2+\dmsol, \; m_3^2=m_1^2+\dmatm$. In this work we use the allowed intervals for the lepton mixing angles, neutrino mass-squared differences and the Dirac phase $\delta$ obtained from the global fit of neutrino oscillation data of ref.~\cite{deSalas:2020pgw}. We will focus on NO since it currently exhibits a slight $2 \sigma$ preference over IO~\cite{deSalas:2020pgw}.

The effective neutrino mass matrix $\Mnu$ of eq.~\eqref{eq:Mnu}, obtained considering the symmetries of the model, must be matched with $\Mnuh$ written in terms of the low-energy parameters. To test the compatibility of our model with the neutrino oscillation data we use a standard chi-squared analysis. The two-texture zero constraints in $\Mnu$, $(\Mnu)_{13}=(\Mnu)_{33}=0$ [see eq.~\eqref{eq:Mnu}], are incorporated in the total chi-squared function, using the Lagrange multiplier method described in ref.~\cite{Alcaide:2018vni}. Our results are presented in figure~\ref{fig:neutrino}. We notice, from the left plot, that the $\mathbb{Z}_8^{e-\mu}$ case for NO is viable at 2$\sigma$ ($\chi^2_\text{min}=4.79$), sharply predicts maximal Dirac CP violation ($\delta \sim 3 \pi/2$) and selects the second octant for the atmospheric mixing angle $\theta_{2 3}$. From the middle plot, we remark the the a lower limit on $m_\text{lightest}$ is imposed $\sim 40$~meV (3$\sigma$), being probed by cosmology, lying above (below) the less (most) conservative Planck bound~\cite{Planck:2018vyg}, and well below the KATRIN limit~\cite{KATRIN:2021uub}. An upper limit on $m_\text{lightest}$, is also predicted, but only at the 2$\sigma$ level for values $\sim60$~meV. Lastly, from the right plot, we notice that the current KamLAND-Zen 800~~\cite{KamLAND-Zen:2022tow} constraint on $m_{\beta\beta}$ disfavours $\mathbb{Z}_8^{e-\mu}$ NO. 

\section{Charged lepton flavour violation}
\label{sec:flavour}
\begin{figure}[!t]
   \centering
   \includegraphics[scale=0.225]{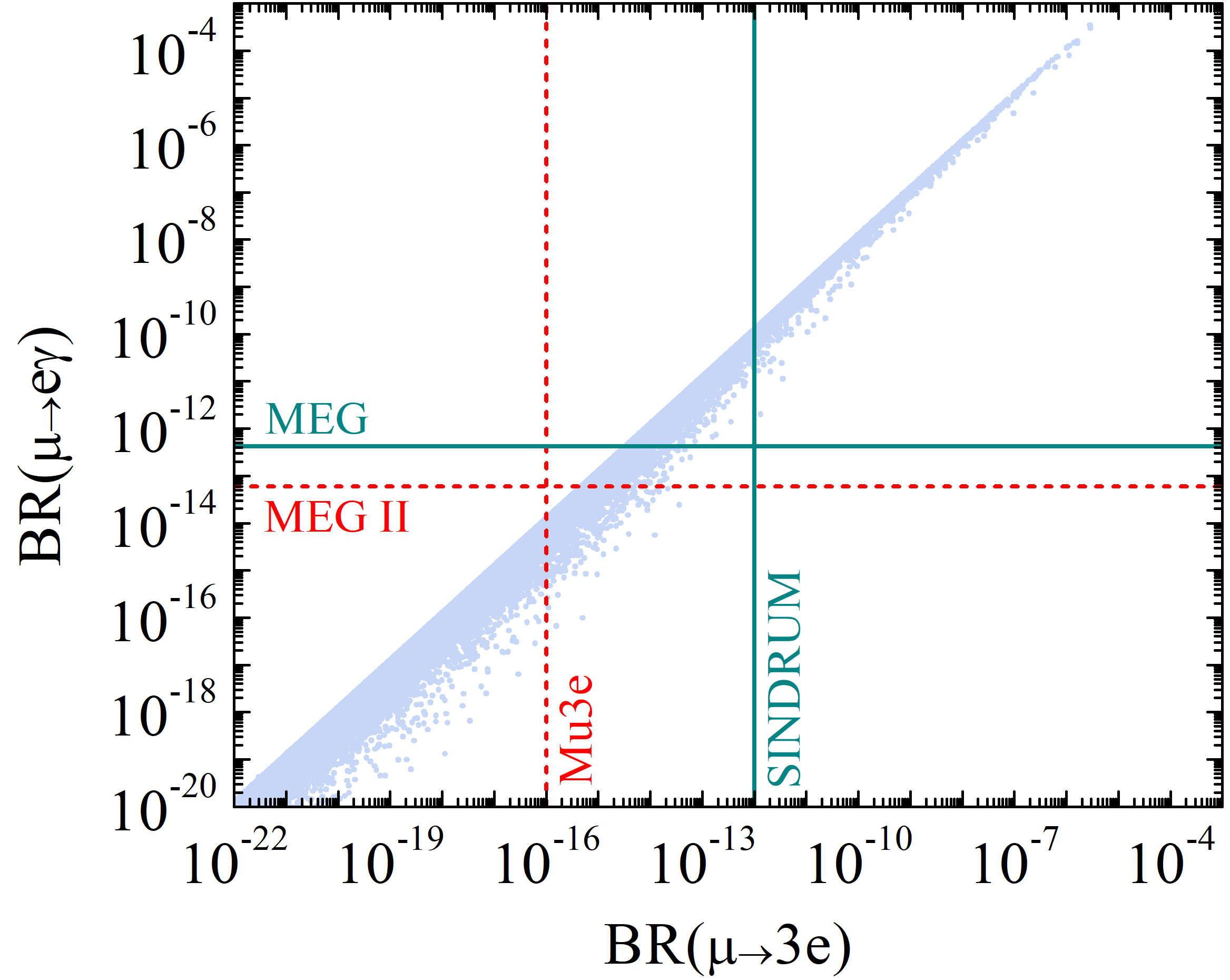} \includegraphics[scale=0.225]{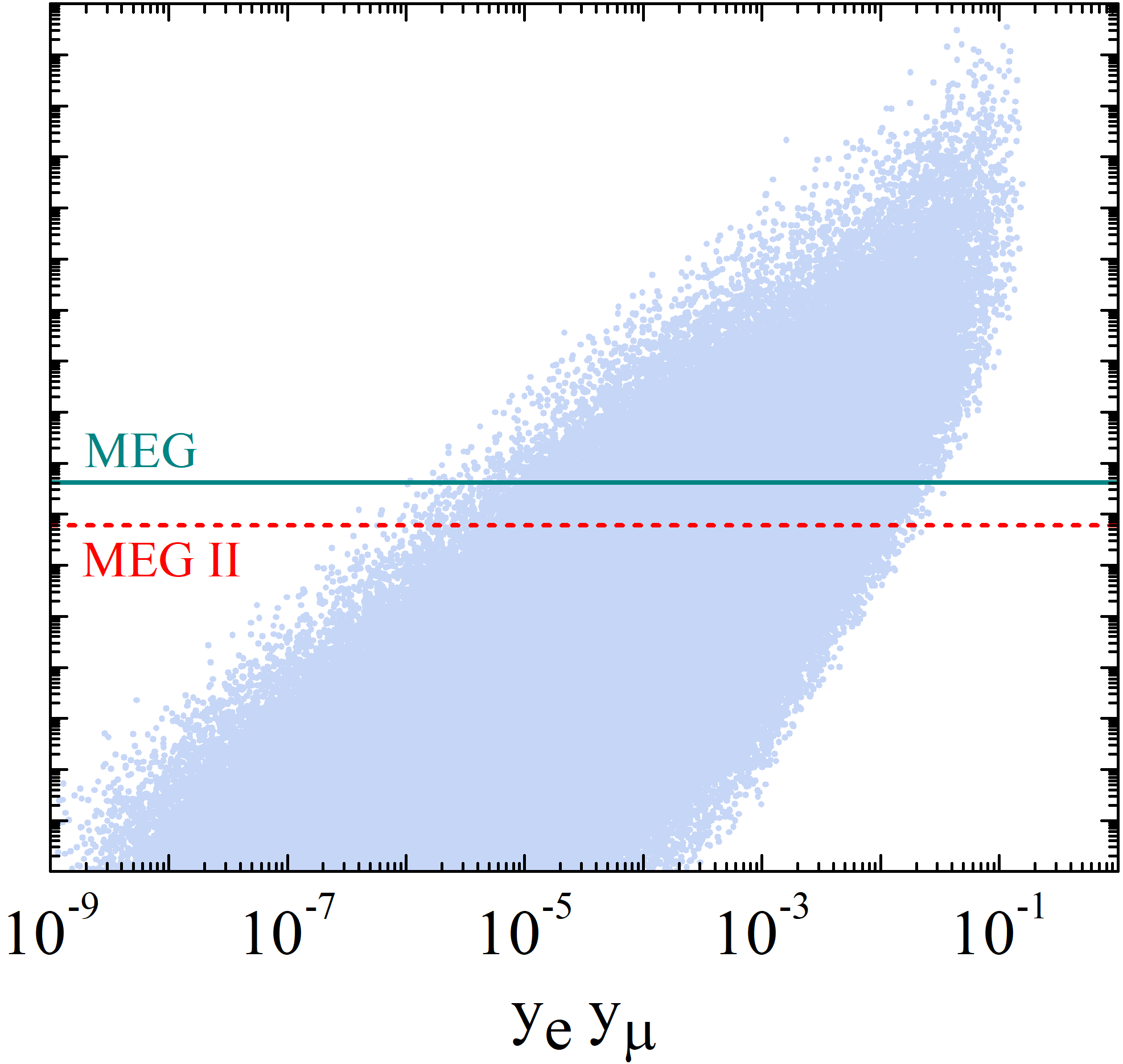} \includegraphics[scale=0.225]{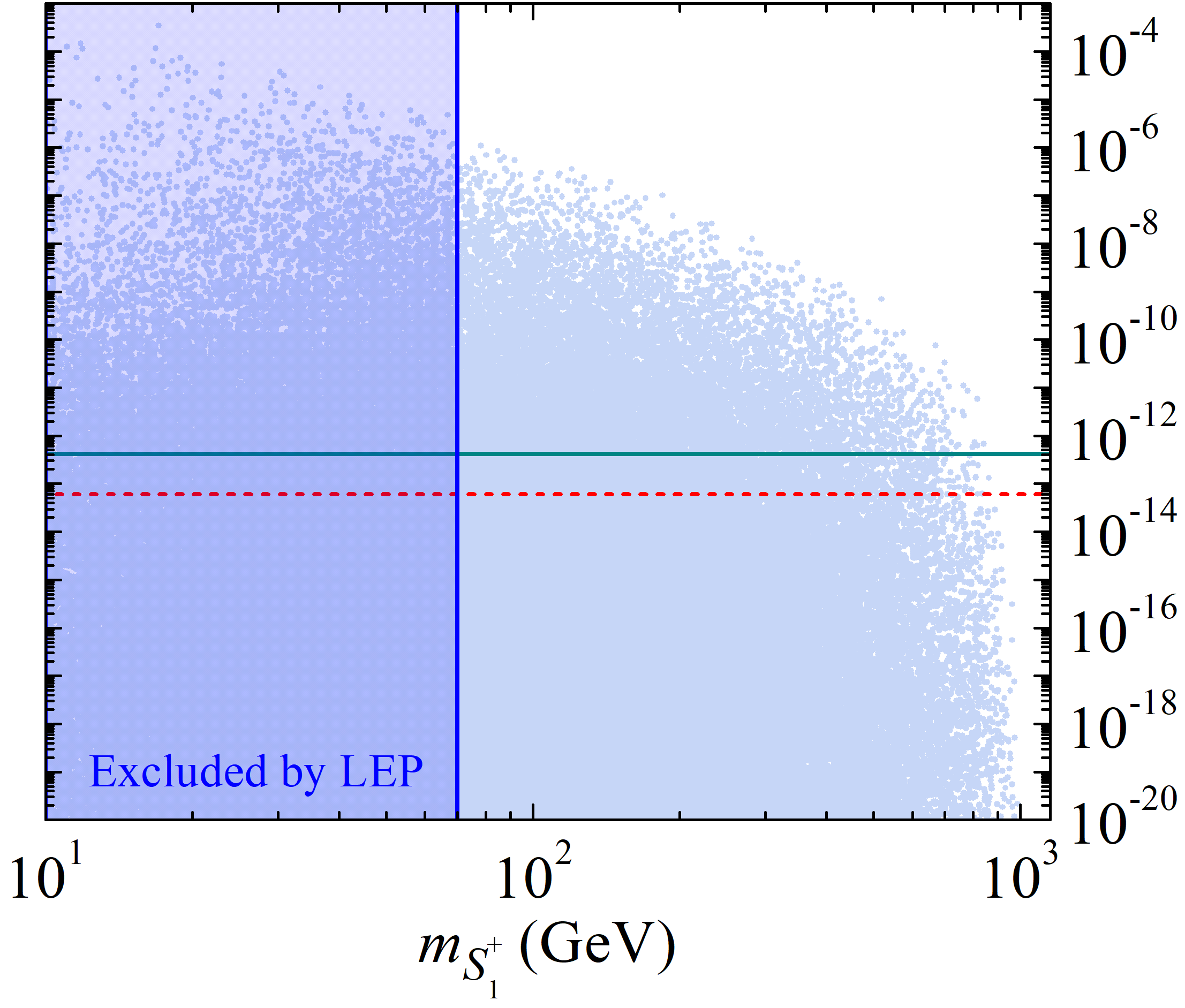}
  \caption{$\BR(\mu \rightarrow e \gamma)$ in terms of $\BR(\mu \rightarrow 3 e)$ (left), the product of Yukawa couplings $y_e y_\mu$ (middle) and lightest \textit{dark}-charged scalar mass $m_{S_1^+}$ (right). The current bounds, MEG~\cite{MEG:2016leq} and SINDRUM~\cite{SINDRUM:1987nra}, and future sensitivities, MEG II~\cite{MEGII:2018kmf} and Mu3e~\cite{Blondel:2013ia}, for the cLFV observables, are respectively indicated by a solid green and dashed red line. The blue shaded-region excludes masses $m_{S_1^+}<70$ GeV.}
  \label{fig:flavour}
\end{figure}
The singly and doubbly charged scalars $\Delta^+$ and $\Delta^{+ +}$ stemming from the Higgs triplet and the \textit{dark}-charged scalars $S_i^+$ ($i=1,2$) from the two inert doublets $\eta_i$ that mix among themselves, will contribute to charged lepton flavour violation~(cLFV). Due to the $\mathbb{Z}_8^{e-\mu}$ flavour symmetry of table~\ref{tab:model}, only four Yukawa couplings to $\Delta$ and $\eta_{1,2}$ are allowed: $(\mathbf{Y}_{\Delta})_{e e} \equiv y_{1}$ and $(\mathbf{Y}_{\Delta})_{\mu \tau} \equiv y_{2}$; $\mathbf{Y}_{f}^1 \equiv y_e$  and $\mathbf{Y}_{f}^2 \equiv y_\mu$, respectively [see eqs.~\eqref{eq:LYuk} and~\eqref{eq:Mnu}]. This provides a very restrictive and highly testable model for cLFV. Namely, the type-II seesaw sector will only contribute to $\tau^- \rightarrow \mu^+ e^- e^-$ mediated by $\Delta^{++}$ at tree-level, while the scotogenic sector leads to  $\mu \rightarrow e \gamma$, $\mu \rightarrow 3 e$ and $\mu - e$ conversion, mediated at one-loop by $S_{1,2}^+$ and $f$. Hence, the triplet and scotogenic contributions never overlap, allowing to distinguish between each neutrino mass generation mechanism. Also, if e.g., $\tau \rightarrow \mu \gamma$ is measured, our $\mathbb{Z}_8^{e-\mu}$ model is ruled out.

In figure~\ref{fig:flavour} we gather our results for muon cLFV~\footnote{The cLFV BRs are computed with \texttt{SPheno}~\cite{Porod:2003um} and \texttt{FlavourKit}~\cite{Porod:2014xia}.}. From the left plot, we notice that relation between the branching ratios~(BR) of $\mu \rightarrow e \gamma$ and $\mu \rightarrow 3 e$ is nearly linear, showing the typical photon-dipole dominance. The spreading in the points is mainly due to the photon-penguin monopole contribution and the mass difference between~$S_1^+$ and~$S_2^+$. We remark that a large fraction of the model's parameter space is excluded by current cLFV bounds (green lines), leaving some regions to be probed by future experiments (dashed red lines). To better understand the middle and right plots, we make use of the analytical expression:
    \begin{align}
    \frac{\BR(\mu \rightarrow e \gamma)}{4.2 \times 10^{-13}} & \approx 1.98 \times 10^{10} \left(\frac{70 \; \text{GeV}}{m_{S_1^+}}\right)^4 \sin^2(2 \varphi) y_e^2 y_{\mu}^2 \left|g\left(\frac{M_f^2}{m_{S_1^+}^2}\right) - \frac{{m_{S_1^+}^2}}{m_{S_2^+}^2} g\left(\frac{M_f^2}{m_{S_2^+}^2}\right)\right|^2,
    \label{eq:BRmueg}
    \end{align}
     where $g(x)$ is the loop function. From the plots and above, it is evident that $\BR(\mu \rightarrow e \gamma)$ depends quadratically on $y_e y_{\mu}$, and the BR value increases for larger $y_e y_{\mu}$ and lower $m_{S^+_1}$. This parameter dependence leaves regions of parameter space above the MEG II projected sensitivity. From the analytical expression we remark that the charged odd-scalar mixing is crucial since if the mixing angle $\varphi$ vanishes there would be no contributions to cLFV from the scotogenic sector.

\section{Dark matter}

\begin{figure}[t!]
    \centering
    \includegraphics[scale=0.135]{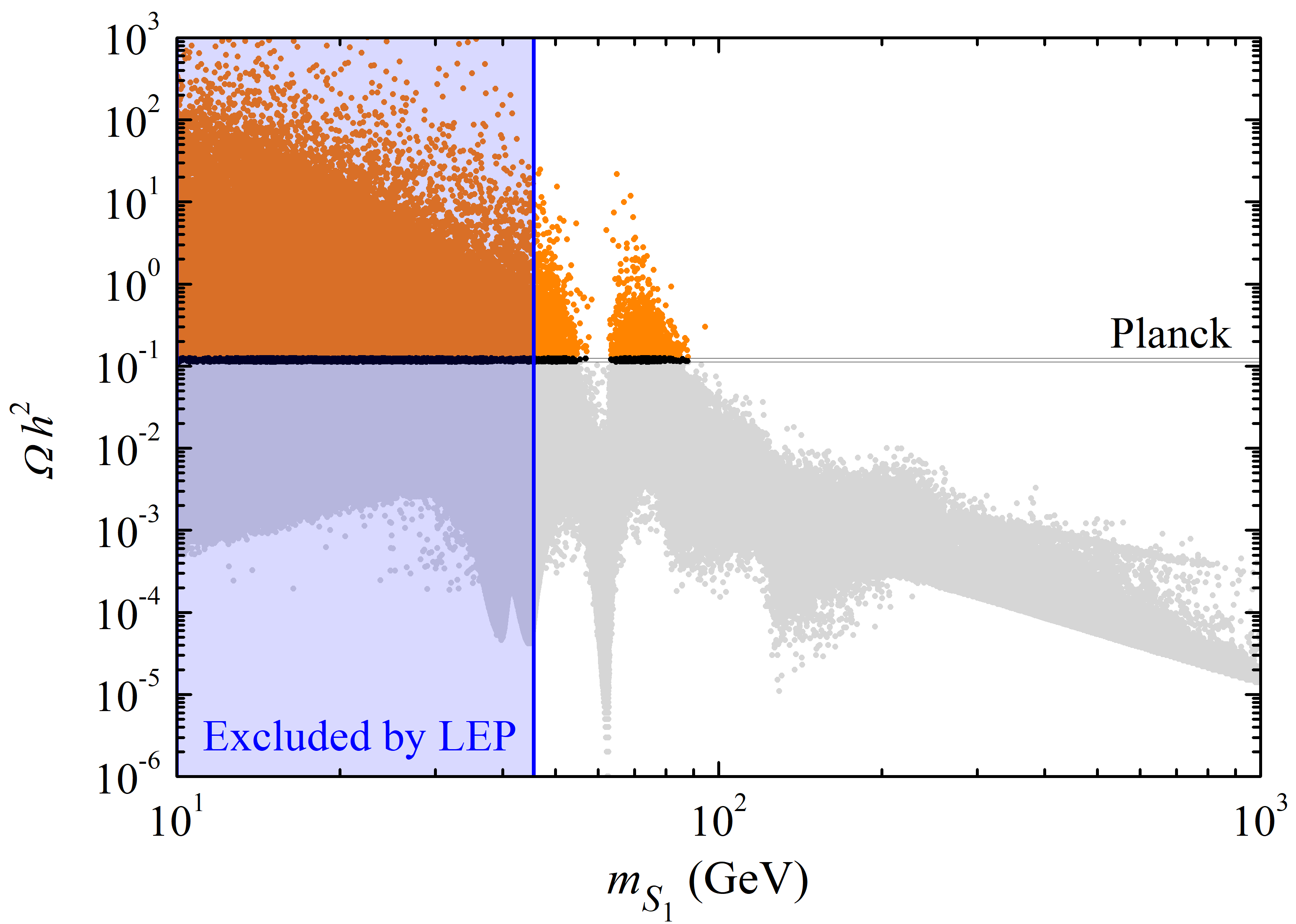} \includegraphics[scale=0.135]{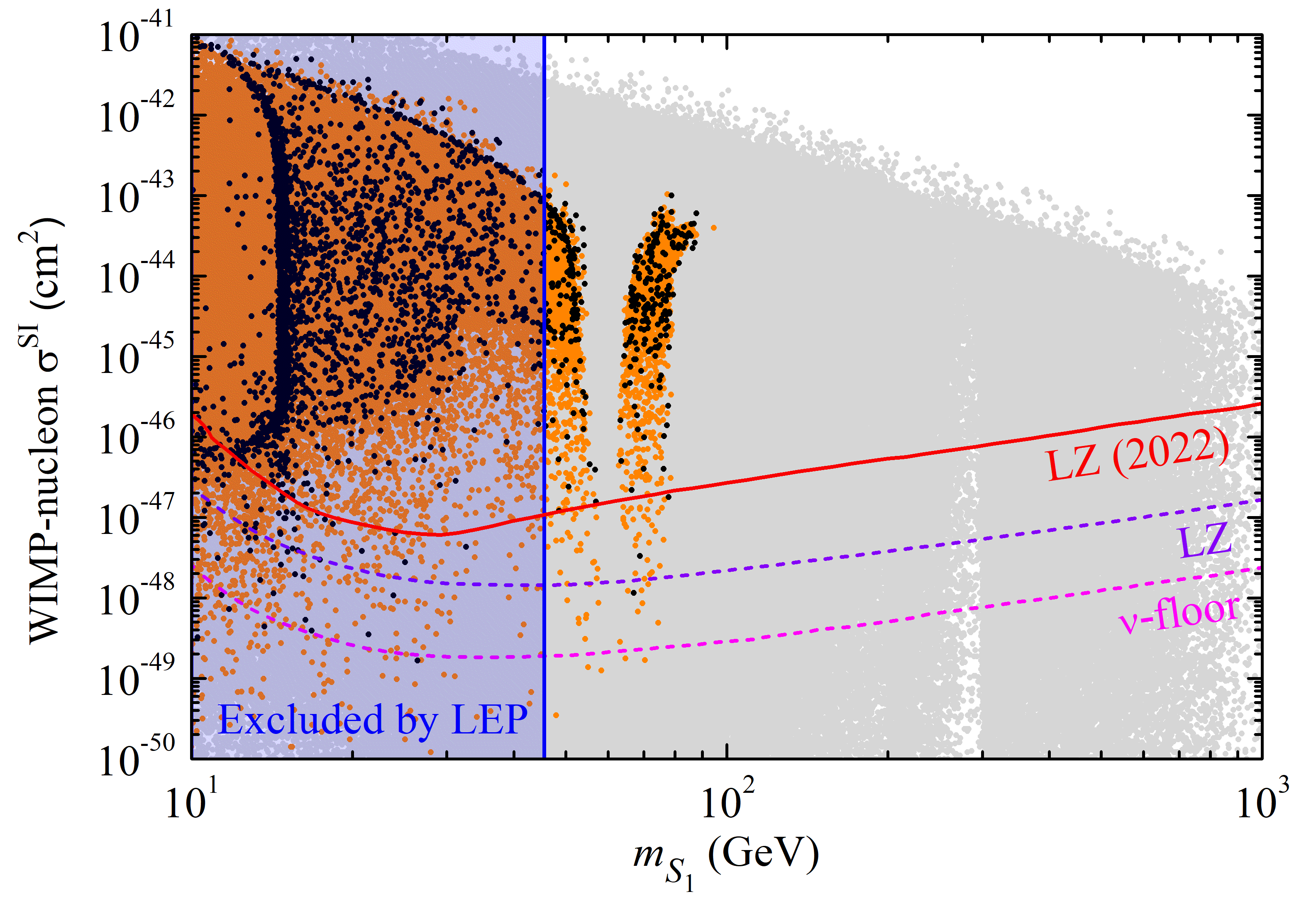}
    \caption{Relic density $\Omega h^2$ (left) and WIMP-nucleon spin-independent elastic scattering cross-section $\sigma^{\text{SI}}$ (right), as a function of the scalar DM mass $m_{S_1}$. Black points lie within the $3 \sigma$ range of CDM relic abundance obtained by Planck~\cite{Planck:2018vyg}. The orange (grey) points give an over (under) abundance of DM. The blue-shaded region is excluded by LEP ($Z$-boson decay width). In the right figure, the solid red line indicates the current bound from the LZ experiment~\cite{LZ:2022ufs}, the dashed violet line represents the future sensitivity for LZ~\cite{LZ:2018qzl} and the dashed magenta line shows the "neutrino floor"~\cite{Billard:2013qya}.}
    \label{fig:DM}
\end{figure}
After SSB, a $\mathbb{Z}_2$ symmetry, under which the \textit{dark}-sector is odd, remains unbroken (see table~\ref{tab:model}). The WIMP DM candidate will be the lightest \textit{dark} particle, namely there are two possibilities: (lightest) neutral scalar~$S_1$ and fermion~$f$. Here we focus on the scalar DM option~\footnote{For a detailed analysis on the fermion DM scenario see ref.~\cite{Barreiros:2022aqu}.}. All generated points are compatible with neutrino oscillation data (see section~\ref{sec:neutrino}) and satisfy the current cLFV bounds (see section~\ref{sec:flavour}). From figure~\ref{fig:DM}, we have~\footnote{The computation of ~$\Omega h^2$ and $\sigma^{\text{SI}}$, is performed with \texttt{MicrOmegas}~\cite{Belanger:2014vza}.}:

\textit{-Relic density}: The black points are within the $3 \sigma$ range for the cold dark matter~(CDM) relic density obtained by Planck~\cite{Planck:2018vyg}, $0.1126 \leq \Omega_{\text{CDM}} h^2 \leq 0.1246$. There are only two viable mass regions that lead to the correct relic density, namely $m_{S_1} \lesssim 60$ GeV and $68 \ \text{GeV} \lesssim m_{S_1} \lesssim 90 \ \text{GeV}$. The lower mass region is excluded up to $m_{S_1} \lsim 45.6$ GeV. The latter bound stems from the LEP precise measurement on the $Z$-boson decay width, which forbids $Z \rightarrow S_1 S_1$. The dips in the relic density, in the left plot, can be understood through the~$S_1$ (co)annihilation channels. The most important feature is that for $m_{S_1} \gtrsim 90 \ \text{GeV}$, our model is not compatible with the Planck $\Omega h^2$ $3 \sigma$ interval, due to a multitude of (co)annihilation channels, e.g. $S_1-S_i$ and $S_1-S_j^+$ for $i=1,\cdots,4$ and $j=1,2$, that lead to a decrease of the $\Omega h^2$ value.

\textit{-Direct detection}: At tree-level the contributions to the spin-independent elastic-scattering cross section $\sigma^{\text{SI}}$ stem from the Higgs $h_1$ and $Z$-boson exchange. From the right figure, we notice that the current constraint on $\sigma^{\text{SI}}$ from LUX-ZEPLIN~(LZ)~\cite{LZ:2022ufs} (solid red line) excludes the viable relic density points for the interval $45.6 \ \text{GeV} \lesssim m_{S_1} \lesssim 60$ GeV. The remaining viable points in the interval $68 \ \text{GeV} \lesssim m_{S_1} \lesssim 90 \ \text{GeV}$ will be probed in the future by LZ~\cite{LZ:2018qzl} (dashed violet line), and other DD experiments up to the "neutrino floor"~\cite{Billard:2013qya}. 

\textit{-LHC Higgs data}: The odd-neutral scalars $S_i$ contribute to the Higgs invisible decay through the channel $h_1 \rightarrow S_i S_j$ ($i,j= 1, \cdots, 4$), if kinematically allowed. The odd-charged scalars $S_i^+$ will contribute at one-loop to $h_1 \rightarrow \gamma \gamma$. The BR for the Higgs invisible decay and diphoton signal strength, are constrained to be, $\BR(h_1 \rightarrow \text{inv}) \leq 0.19$ and $R_{\gamma \gamma} = 1.11^{+0.10}_{-0.09}$, respectively~\cite{ParticleDataGroup:2020ssz}. It can be shown that these LHC Higgs data constraints also exclude the low-mass region $45.6 \; \text{GeV} \lesssim m_{S_1} \lesssim 60 \; \text{GeV}$ (see ref.~\cite{Barreiros:2022aqu} for details). Thus, we conclude that the intermediate mass region $68 \ \text{GeV} \lesssim m_{S_1} \lesssim 90 \ \text{GeV}$ for scalar DM is viable since it leads to a correct relic density while evading current collider and DD experimental constraints.

\section{Conclusion}

We have studied the phenomenology of a hybrid scoto/type-II seesaw model which, albeit being based on a very simple flavour symmetry, together with the feature of SCPV, leads to significant constraints in view of present and future data coming from neutrino oscillation, neutrinoless double beta decay, cLFV and DM experiments.

\acknowledgments

This work is supported by Fundação para a Ciência e a Tecnologia (FCT, Portugal) through the projects CFTP-FCT Unit UIDB/00777/2020 and UIDP/00777/2020, CERN/FIS-PAR/0004/2019, partially funded through POCTI (FEDER), COMPETE, QREN and EU. The work of D.B. and H.C. is supported by the PhD FCT grants SFRH/BD/137127/2018 and 2021.06340.BD, respectively.

\end{document}